# Advanced engineering design as practiced today from the view point of the CERN Industrial Liaison Officer

Michele Barone
Institute of Nuclear Physics
NCSR"DEMOKRITOS"-Athens-Greece
E-mail1:barone@inp.demokritos.gr
E-mail2:michele.barone@cern.ch

## Abstract

After an introduction about CERN,a brief description of the Large
Hadron Collider(LHC) it is reviewed .Pros and cons of a few advanced engineering design cases are taken in consideration together with the involvement of the European Industry.The conclusion is that the LHC project has been an important driving force for Innovation in European Industry.

## I.Introduction

### 1.1 Introduction to CERN

CERN, the European Organization for Nuclear Research, is an intergovernmental organization with 20 Member States.

Its seat is in Geneva, but its premises are located on both sides of the French-Swiss border. CERN's mission is to push back the frontiers of knowledge(e.g. the secrets of the Big Bang …what was the matter like within the first moments of the Universe's existence?), to develop new technologies for accelerators and detectors, Information Technology (the Web and the GRID,Medicine - diagnosis and therapy),to train scientists and engineers of tomorrow. To enable international collaboration in the field of high-energy particle physics research , it designs, builds and operates particle accelerators and the associated experimental areas. At present more than 10.000 scientific users from research institutes all over the world are using CERN's installations for their experiments.

The accelerator complex at CERN is a succession of machines with increasingly higher energies, injecting a particle beam each time into the next accelerator, bringing the beam to an energy increasingly higher. Protons are obtained by removing electrons from hydrogen atoms, then they are injected from the linear accelerator(LINAC2) into the PS Booster,then to the Proton Synchrotron(PS),followed by the Super Proton Synchrotron(SPS),before reaching the Large Hadron Collider at the energy of 450GeV. Protons circulate in the LHC for 20 minutes before to reach the maximum energy of 7 TeV/beam.

Lead ions for the LHC start from a source of vaporized lead and enter LINAC3 before being collected and accelerated into the Low Energy Ion Ring (REIR).They then follow the same route to maximum acceleration as the protons.

The flagship of the complex is the Large Hadron Collider (LHC) as presented below:

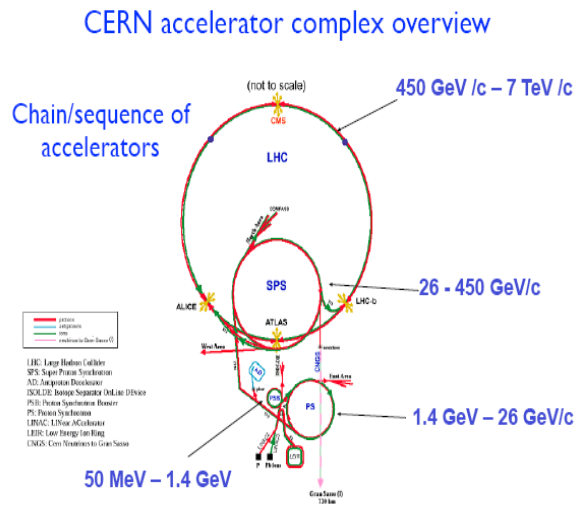

Figure 1 Accelerators complex

## 2.The Large Hadron Collider (LHC)

The LHC is based on 1232 double aperture superconducting dipole magnets,equivalent to 2864 single dipoles, operating up to 9 Tesla.The machine also incorporates about 500 "two-in-one"superconducting quadrupole magnets with a gradiend of more 250T/m. Moreover there are more than 400s.c.corrector magnets of many types.In operation the machine is involving the cooling down to 1.9K of 40.000 tonnes of material.The LHC is installed in the exhisting 27 km circumference tunnel ,about 100 m underground,previously housing the Large Electron Positron Collider(LEP).The beam stored energy is 36MJoules.
The beams are crossing in 4 points around which very large particle detectors are built. The starting point of the project is considered March 1984,but it was approved in 2000 and inaugurated in September 2008.Its final costs has been 4.6 billion of swiss francs, including the contribution from States not Members of the Organization.It has been a challenging project from technical point of view and has proved to be an enriching experience for the participants.

**3.Some Technological Issues:Pros**

The LHC construction , because of the approved budget, had to make the maximum use of the exhisting infrastructure in order to reduce the total cost.This implied some constraints on the technical point of view: since the tunnel hosting the new accelerator was the old 27km of the LEP era,it was not possible to modify it therefore the equipments had to be mounted in its 3.8m diameter.

A s.c.magnet occupies a considerable amount of space because of its cryostat, therefore it was impossible to fit in the same tunnel two independent rings as in the case of the Superconducting Super Collider proposed in USA.as shown in fig2 :

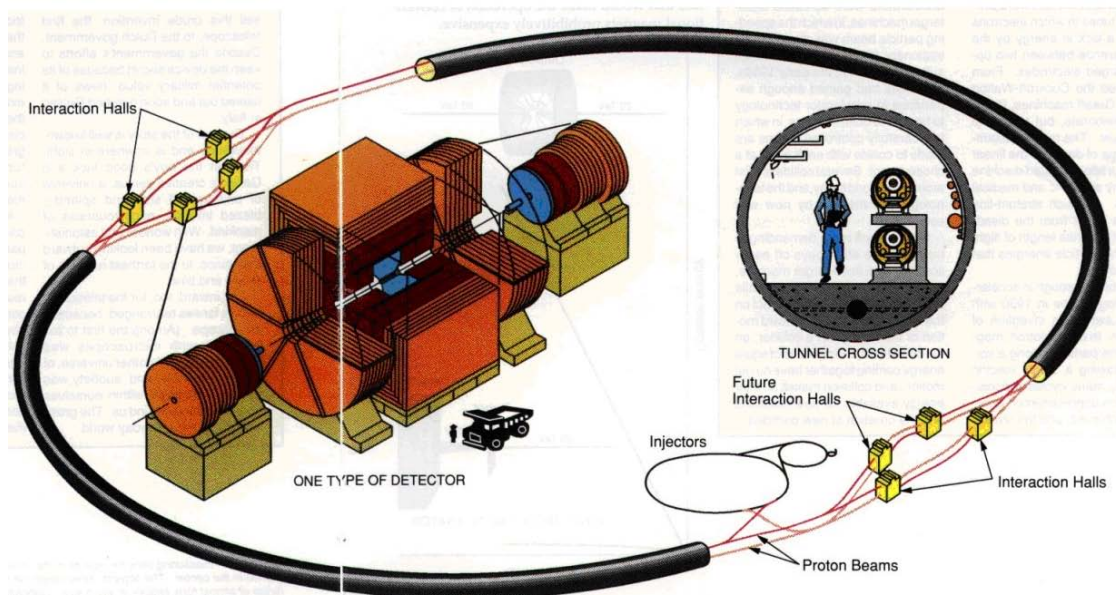

Figure 2 –Schematic View of the SSC

A novel design with two rings separated by only19cm inside a common yoke and cryostat was developed: it was called the"two-in –one'approach.
The fig3 shows a detailed drawing of a typical element:

# Cross-section of the cryodipole

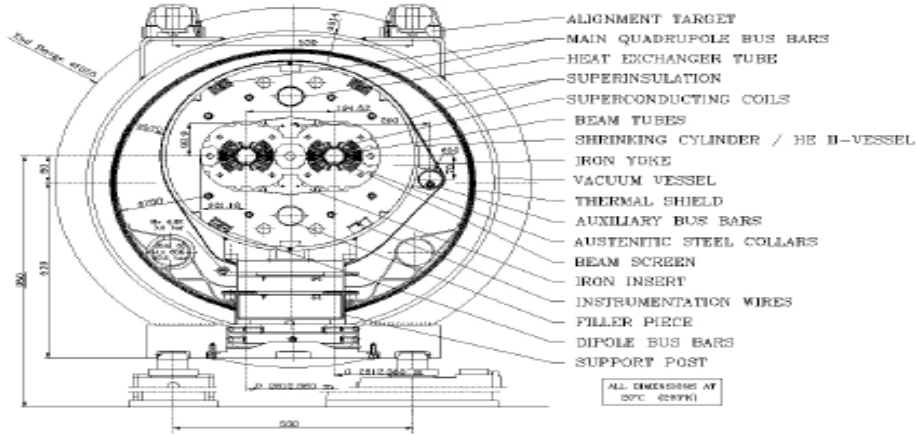

Figure 3  -Cross section of cryodipole

The sensitivity of the forces acting on the cold masses (because of the changes during the cool-down) to the tolerances on all major components  like
collars laminations,inserts etc.,has been checked by Finite Element Analysis computation by applying statistical methods infact about 3000 geometries have been computed.Just to mention some figures ,the forces acting on the coils are of 15 MPa and to avoid any distortion of them which could trigger a
quench ,12 millions of non –magnetic collars  with a tolerances from+/-0.02 to 0.03mm have been produced by the European Industry.The total dipole length is 15m.This version,has been a real challenge and it generated   a saving of 10% of the total  project cost(4.6Gchf).
The  involvement of the Industry   since the beginning of the project in 1984 has produced good results.Three companies have been in charge of the construction of the LHC's s.c.dipoles:the French consortium AlstomMSA-Jeumont,the Italian firm Ansaldo Superconduttori and the German company Babcook Noell Nuclear.Each has provided one-third of the 1248 magnets including a pre-serie .
A good uniformity in the dipoles mass production was found and one manufacturer delivered 1 year in advance.The figure 4 shows a tipical dipole having a weight of 30 tons.

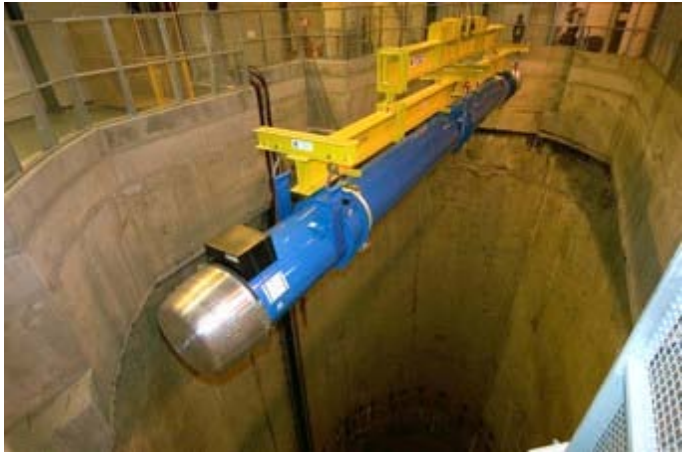

Figure 4 - Lowering of dipole into the tunnel.

Another example of Advanced Engineering case is given by the Compact Muon Solenoid (CMS) detector at point 5.
The experimental program for the LHC began in March 1992 with a meeting in Evian-les Bains ,near the French part of Lake of Geneva during which,the CMS Collaboration presented an Expression of Interest.The construction of it has presented formidable challenges from the technological ,engineering ,organizational and financial point of view.The construction has also required the pooling of the resources and talent of a large number of people as well as the involvement of the World Industry.CMS is formed by more than 2500 scientists and engineers from 180 Institutions in 38 countries around the world.

The CMS detector is inspired by the LEP experiment,it has an onion-like structure with a succession of detectors both in the barrel and in the forward regions to determining the event topology, the nature and the energy of emerging particles.The momentum of the charged tracks is derived from the curvature of a strong magnetic field,the particle identification is made by a succession of electromagnetic and hadronic calorimeter followed by a muon detector.The fig 5.is a schematic of the slice of the transverse cut trough CMS ,illustrating the identification of the particles produced in the proton-proton collisions.

# Particle interactions in detectors

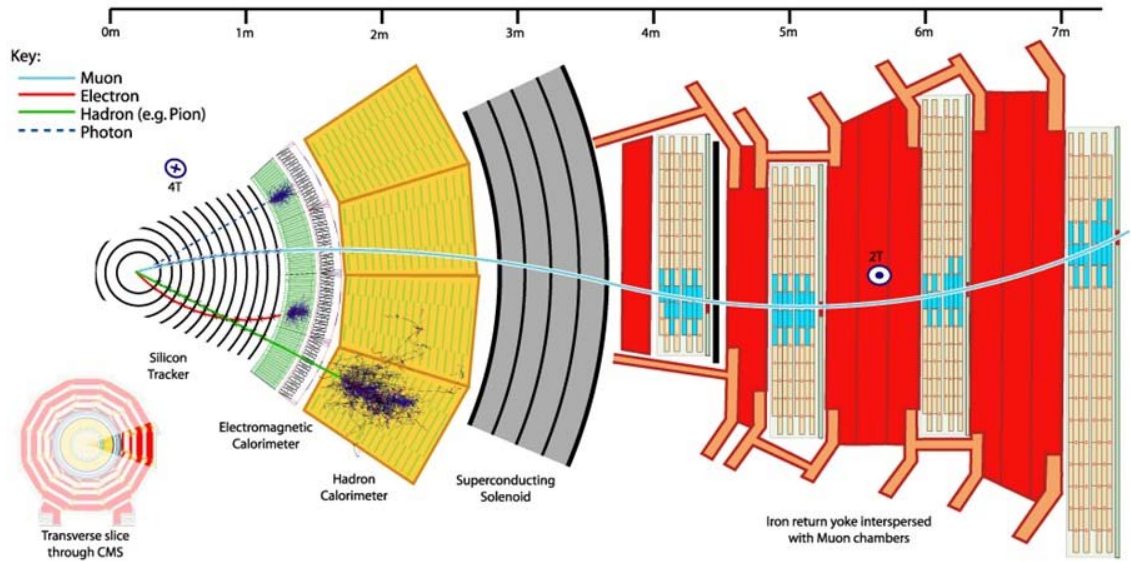

Figure 5 Slice of transverse cut trough CMS

More precisely,the magnet system consists of a 4 Tesla Solenoid Superconducting Coil, having a 6m free bore,13m length,enclosed in a return Yoke,comprising the Barrel Yoke and the two End-CapYokes.The return Yoke is designed as a regular twelve-side structure.The main dimensions of the complete detector are:length 21m,outer diameter14.8m and a total mass 12.500 tons. Remarkable challenges were faced during the construction of this massive detector: it was built in modules and subsequently assembled inside a cavern 100m below the surface.Already at the conceptual design stage 20 years ago,it was decided to divide the massive flux-return iron yoke of the solenoid in sections allowing the construction and the assembly in stages.The fig 6 shows the detector sectioned in its parts.

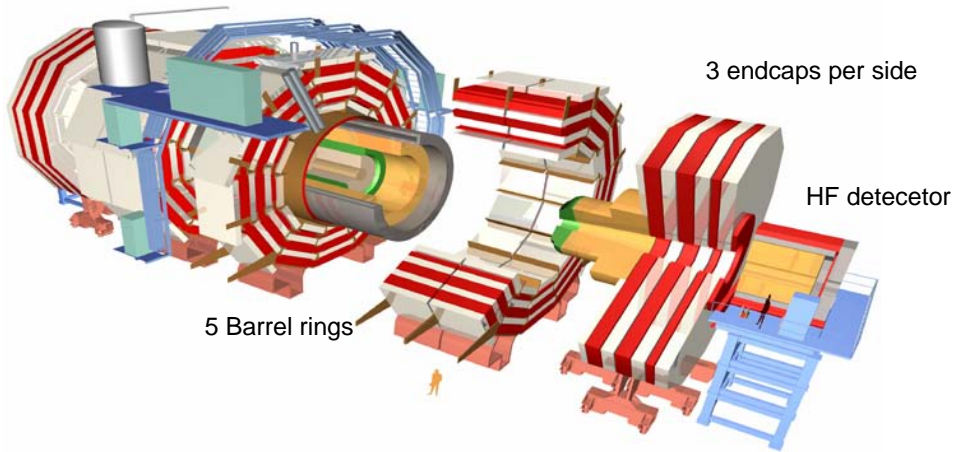

Figure 6 –CMS detector in slices

The yoke was assembled in a large surface building specially conceived for this purpose and equipped with a large gantry crane constructed on the outside of the building to lower the heavy elements of the detector in 15 pieces. See Figure 7

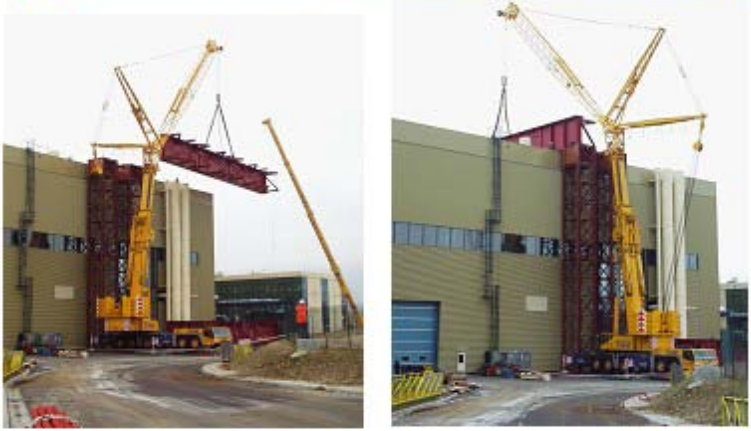

Figure 7 CMS gantry installation

The most heavy slice had a weight of 1920 tons.For the lowering operation,elements were suspended by four massive cables each with 55 strands,and attached to a step-by step hydraulic jacking system.Each operation took around 10 hours.The fig8,shows the lowering of the final element.

## Lowering of the final element of CMS into its cavern

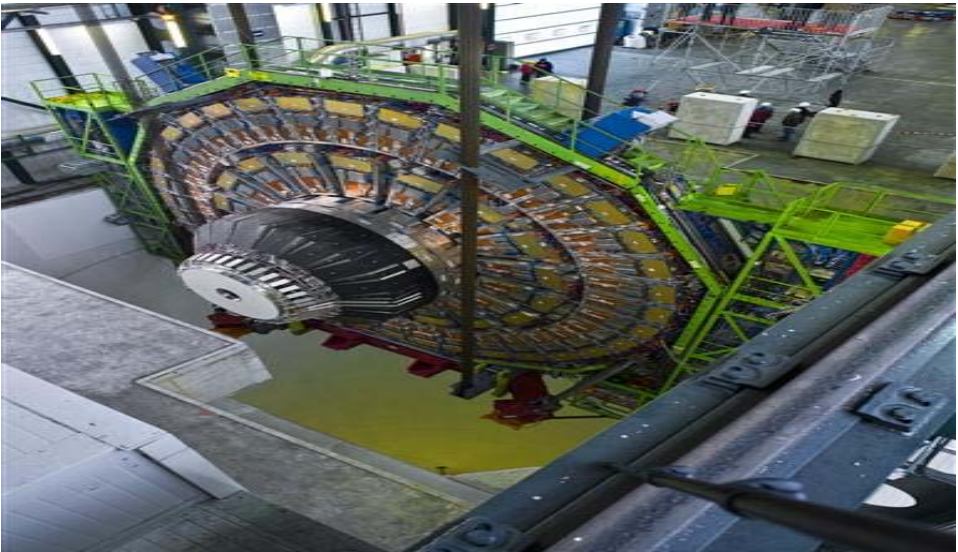

Figure 8 –Lowering of the final element into the CMS cavern

Many have been the advantages in planning an experiment in such way:time was saved by working simultaneously on the detector while the experimental cavern was being excavated,fewer risks working at the surface in a more spacious area,each element and all elements togheter could be tested before lowering.When safely positioned in the cavern all the instrumentation was connected to the service sources.

**4.LHC design flaws**

On 19 September 2008, just weeks before the LHC was first scheduled to start colliding protons, an electrical short caused massive damage. A connection between two superconducting cables developed a small amount of resistance, which warmed the connection (busbar splices) until the cables — cooled by liquid helium to superconducting temperatures — lost their ability to carry current. Thousands of amps arced through the machine, blowing a hole in its side and releasing several tonnes of liquid helium. The expanding helium gas created havoc, spewing soot into the machine's ultraclean beamline and ripping magnets from their stands. Repairs took more than a year, and the LHC successfully restarted last November 2009.An investigation revealed that technicians had not properly soldered the cables together. With tens of thousands of such connections, it is perhaps inevitable that some were faulty, but design flaws worsened the problem. The silver–tin solder that was used melted at high temperatures and did not flow easily into the cable joints. Moreover, workers did not adequately check to see if each connection was electrically secure. Sensors to detect an overheating circuit, which might have helped prevent the accident, were not installed until after it happened.In addition  when the wires were originally joined, the same silver–tin solder was used to connect them to an adjacent copper stabilizer, meant to provide an escape route for current in the event of a failure. That step risked reheating and destroying the original connection. Making the second connection to the stabilizer with a different type of solder that had a lower melting point could have avoided the problem. , souch solution was considered and rejected because the alternative solder contained lead, a hazard to workers. Figure 9shows a tipical busbar splice .

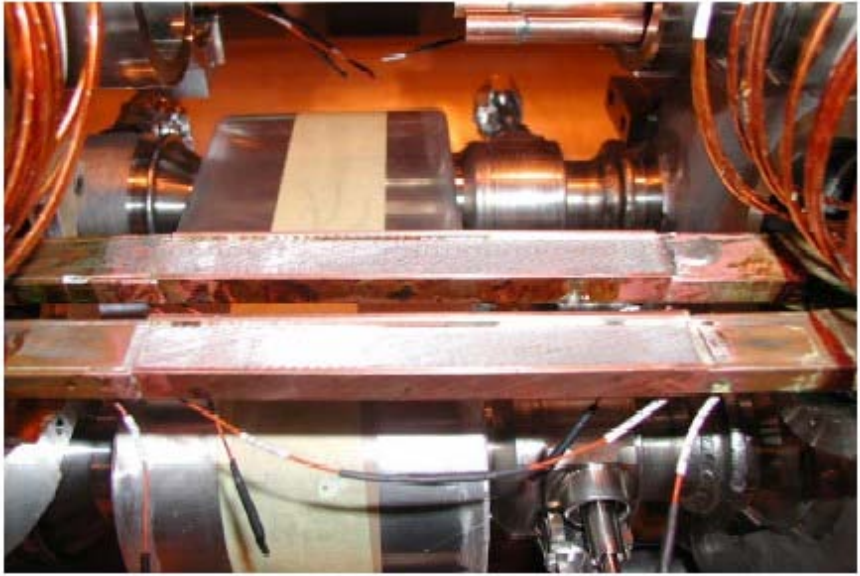

Figure 9 busbar splices

In a paper published on 22 February 2010 (L. Rossi Supercond. Sci. Technol. **23, 034001; 2010**), there is the conclusion that the catastrophic failure of a splice between two magnets was not a freak accident but the result of poor design and lack of quality assurance and diagnostics.

**Conclusion**
 In spite the flaws, the LHC project has been an important driving force for Innovation in European Industry.

**Acknowledgements**

The author whishes to express his sincere thanks to prof.C.Spitas and his group at  Delft University Technology for the invitation to the Symposium on "Case studies in advanced engineering design" and for the hospitality provided.